# Analysis of the volatiles in the headspace above the plasmodium and sporangia of the slime mould (*Physarum polycephalum*) by SPME-GCMS

**Huda al Kateb[1] and Ben de Lacy Costello[1]**

**[1]Institute for biosensing technology, University of the West of England, Bristol, BS161QY, UK**

E-mail: Ben.DeLacyCostello@uwe.ac.uk

## Abstract

Solid phase micro-extraction (SPME) coupled with Gas Chromatography Mass Spectrometry (GC-MS) was used to extract and analyse the volatiles in the headspace above the plasmodial and sporulating stages of the slime mould *Physarum Polycephalum*. In total 115 compounds were identified from across a broad range of chemical classes. Although more (87) volatile organic compounds (VOCs) were identified when using a higher incubation temperature of 75$^{o}$C, a large number of compounds (79) were still identified at the lower extraction temperature of 30$^{o}$C and where the plasmodial stage was living. Far fewer compounds were extracted after sporulation at the two extraction temperatures. There were some marked differences between the VOCs identified in the plasmodial stage and after sporulation. In particular the nitrogen containing compounds acetonitrile, pyrrole, 2, 5-dimethyl-pyrazine and trimethyl pyrazine seemed to be associated with the sporulating stage. There were many compounds associated predominantly with the plasmodial stage including a number of furans and alkanes.

Interestingly, a number of known fungal metabolites were identified including 1-octen-3-ol, 3-octanone, 1-octen-3-one, 3-octanol. In addition known metabolites of cyanobacteria and actinobacteria in particular geosmin was identified in the headspace. Volatile metabolites that had previously been identified as having a positive chemotactic response to the plasmodial stage of *P. polycephalum* were also identified including β-farnesene, β-myrcene, limonene and 3-octanone.

This study constitutes the first comprehensive analysis of the headspace volatiles emitted from *Physarum Polycephalum*. Further work to understand the origin and function of the volatiles identified is required.

## 1. Introduction

*Physarum polycephalum* is a member of the supergroup *Amoebozoa* [Baldauf *et al*. 2000] and of the class/superclass *Myxogastridae* (or *myxomycetes*) commonly referred to as plasmodial or true slime molds. Although historically classified as fungi, molecular data now clearly show that they are most closely related to the cellular slime molds (*Dictyostelidae*). The vegetative stage (macroplasmodium) is a large, single cell containing multiple dipoid nuclei that divide precisely at the same time. Macroplasmodia migrate by protoplasmic streaming which reverses *circa* every 60 s. [Kishimoto 1958] Plasmodia engulf bacteria, *myxomycete* amoebae and other microbes [Dussutour 2010]. They also secrete enzymes for digesting the engulfed material. Starvation of the macroplasmodium in the presence of light induces differentiation into specialized sacs (sporangia), a highly regulated process, which includes the complete conversion of the macroplasmodium into "fruiting bodies". Haploid spores are produced inside the fructification by meiosis. The germinated spore can transform into either an amoeba-like myxamoeba cell or a flagellated swarm cell (mxyoflagellate) [Sauer 1982]. It has been shown that in addition to the conditions mentioned both calcium and malate are sporulating-promoting factors [Renzel *et al.* 2000].

Cytoplasm streams rhythmically through a network of tube, circulating nutrients and chemical signals and forming pseudopods that allow the organism to navigate and respond to its environment. Under adequate nutrition, *P. polycephalum* plasmodia are completely sedentary and grow steadily [Ashworth and Dee 1975], but on nonnutrient substrates, they migrate a few centimeters per hour [Halvorsrud and Wagner 1998], directed by external stimuli, including gradients of nutrients such as sugars and proteins [Carlile 1970, Ueda 1975 and Durham and Ridgway 1976]. When two or more identical food sources are presented at various positions to a starved plasmodium, it optimizes the shape of the network to facilitate effective absorption of nutrients [Nagaki *et al*.2000], and plasmodia select the higher concentration patch when two patches differ in nutrient concentration [Dussutour 2010].The plasmodium not only propagates according to the position of nutrients but also in response to external/environmental gradients in light level and humidity. It is also well established that *P. polycephalum* will propagate according to gradients in certain chemical species, either chemoattractants or chemorepellents. The *P. polycephalum* plasmodium is a model system for studying non-muscular motility, and its chemotactic behavior has been well documented [Durham and Ridgway 1976, Kincaid and Mansour 1978ab, Ueda and Kobatake 1982, Beylina *et al.* 1996] In particular, substances causing negative taxis (repellents) were shown to increase the period of contractility and to decrease the area of spreading when present uniformly within the substrate. [Ueda and Kobatake, 1982 and Beylina *et al.* 1996]. Experimental studies confirmed that the following substances acted as chemoattractants for the plasmodium, glucose, galactose, maltose and mannose, [Carlile 1970, Knowles and Carlile 1978] peptones [Carlile 1970, Coman 1940] the amino acids phenylalanine, leucine, serine, asparagine, glycine, alanine, aspartate, glutamate; and threonine,[Chet et al. 1977, Kincaid and Mansour 1978b, and McClory and Coote 1985] phosphates, pyrophosphates, ATP and c AMP and thorium nitrate. [Ueda *et al.* 1976] A plasmodium is allegedly indifferent to fructose and ribose.[Carlile 1970, Knowels and Carlile 1978] Whereas, the following compounds have been found to act as chemorepellent molecules, sucrose and inorganic salts such as the chloride salts of (K, Na, $NH_4$, Ca, Mg, La)[Ueda *et al.* 1976, Adamatzky 2010] and tryptophan.[McColry and Coote 1985] Therefore, it is clear that the nutritional value of the substance is not paramount in determining either chemoattractant or chemorepellent properties.[Kincaid and Mansour 1978a,b] For some substances, the effect on the plasmodium can be determined by the proximity of the organism to the source (or the concentration of the source), meaning that some substances can act as both chemoattractant and chemorepellent molecules. An example is the sugars galactose and mannose, which are reported to act as chemoattractants[Carlile 1970, Knowels and Carlile 1978] and chemorepellents that inhibit motion.[Denbo and Miller 1978]

Recently it was found [Adamatzky 2011] that the plasmodium is strongly attracted to herbal calming/somnipherous tablets Nytol and KalmsSleep. To select the principle chemoattractant in the tablets, laboratory, experiments were undertaken on the plas-modium's binary choice between samples of dried herbs/roots: *Valeriana officinalis*,

*Humulus lupulus, Passiflora incarnate, Lactuca virosa*, *Gentiana lutea* and *Verbena officinalis*. A hierarchy of chemo-attractive force was calculated from the binary interactions and it was found that Valerian root was the strongest chemo-attractant for *P. polycephalum* of the substances tested. Valerian contains hundreds of identified, and possibly the same amount of not-yet-identified components, including alkaloids, volatile oils, valerinol and actinidine. Therefore, it is unclear which component is causing the chemo-attractive effect. However, actinidine is known to have a chemo-attractive effect on a number of other animal species so remains a strong candidate [Adamatzky and de Lacy Costello 2012]

In previous work [de Lacy Costello and Adamatzky 2013] we found that *P. polycephalum* exhibited positive and negative chemotaxis to a range of volatile organic compounds. The chemoattractive compounds in order of strength of action were as follows: Farnesene > β-myrcene > tridecane > limonene > p-cymene > 3-octanone > β-pinene > m-cresol > benzylacetate > cis-3-hexenylacetate. The chemo repellent compounds in order of strength of action were: nonanal > benzaldehyde > methylbenzoate > linalool > methyl-p-benzoquinone > eugenol > benzyl alcohol > geraniol > 2-phenylethanol.

Kincaid and Mansour [Kincaid and Mansour 1979] found that inhibitors of the enzyme cyclic 3',5'-AMP phosphodiesterase act as strong or moderate chemoattractants in P. polycephalum. Among the substances tested the strongest effect was observed with 4-(-3-butoxy- 4-methoxybenzyl)-2-imidazolidinone and moderate effects from theophylline and other xanthine derivatives (interestingly they observed negative chemotaxis at high concentrations).

There has been much work carried out studying the volatile secretions of fungi and bacteria [Bennett *et al.* 2012, Morath *et al.*2012]. Most of this has been motivated by the desire to detect disease in humans [Garner et al. 2008], spoilage of foodstuffs [Schnurer et al. 1999], environmental exposure e.g indoor air quality, sick building syndrome [Straus et al. 2003, Cooley et al. 1998] etc. rather than to understand the metabolic processes *per se*. However, some studies have focussed on the understanding of metabolism in order to exploit biotechnological production e.g. for biofuels, biocontrol, food flavouring and aroma compounds etc. [Morath *et al*. 2012, Abrahoa *et al.* 2013]. There has also been a great deal of work undertaken in analysing the volatile secretions of plants [Kesselmeier and Staudt 1999] and insects [Metcalf 1987]. This again has a predominantly commercial motivation for developing insect repellents, crop protection, flavour products etc. However, there are some studies which investigate the evolution of common signalling molecules within ecosystems [Schiestl 2010]. Slime molds play an important role within the temperate and tropical forest ecosystems where they are important heterotrophs in the decomposition of organic matter. Therefore, it is probable that their secretion and response to VOCs would have evolved in partnership with these other organisms.

However, to date no studies have analysed the volatiles emitted from *P. polycephalum*. The aim of the study was to assess and investigate the headspace volatiles above the plasmodial phase and post sporulation stage of *P. polycephalum*.

## 2. Materials and Methods

**2.1. Culturing of *Physarum polycephalum*.** The true slime mold, the plasmodium of *Physarum polycephalum* (strain HU554 × HU560), was cultured with oat flakes on a 1% non-nutrient agar gel at 25°C in the dark. Both plasmodia and spores of the slime mould (*Physarum polycephalum*) were extracted without media (0.5g) and placed in a 10 ml glass vial (Supelco, Poole, UK) sealed with a PTFE/silicone rubber septum. The volatiles were extracted immediately using the method described in section 2.2.

## 2.2. Headspace volatile collection by Solid Phase Micro-extraction

The volatiles were sampled statically by the exposure of the SPME fibre to the headspace above the cultures. Sampling of the headspace volatiles was undertaken using 75 µm carboxen/ polydimethylsiloxane (CAR/PDMS) fibre (Sigma-Aldrich, Dorset, UK). The fibre was thermally conditioned in the injection port of a GC (Clarus 500, Perkin Elmer, Beaconsfield, UK) at 250°C for 10 minutes prior to each analysis, and a blank run of the fibre and the GC column were carried out after conditioning to ensure both were clean.. For the volatile extraction the vials were placed in a heating block (TECHNE Dry-Block® DB-3, UK) and were equilibrated at 75°C and 30°C for 5 mins prior to the volatile analysis. The volatile collection was carried out at 75°C for 1hr and at 30°C for 3h. The analysis was repeated with new cultures at both extraction temperatures. Empty vials were extracted at both temperatures to serve as a control.

## 2.3. Gas chromatography mass-spectrometry (GC-MS) analysis

Analyses were carried out using a Perkin Elmer Clarus 500 GC-MS (PerkinElmer, Beaconsfield, UK). The capillary column used was a 60 m long Zebron ZB-624 capillary GC column with an inner diameter of 0.32 mm I. D. and 1.4 µm film thickness (Phenomenex, Macclesfield, UK). The stationary phase consisted of 94% dimethyl polysiloxane and 6% cyanopropyl-phenyl. The GC was operated in splitless mode. The GC injection port was fitted with a 1 mm quartz liner and heated to 250°C. The carrier gas was helium (BOC, Guildford, UK) with a flow of 1.1 ml min-1.

The GC was held at 35°C for one min and then ramped up by 7°C min-1 to 100°C with no hold, then 4°C min-1 to 200°C with no hold, then at a rate of 25°C min-1 the temperature was raised to 250°C with a 38 min hold giving a total run time of 75.29 min. MS scan time was from 0-75.29 min, with no solvent delay. The MS system was operated in the electron ionization (EI) mode at 70eV, scanning the mass ion range 10-300 with a scan time of 0.3 seconds and interscan delay of 0.05 seconds.

### 2.4. Volatile Identification

The volatile identification was determined using Turbomass 5.4.2 and the NIST 2005 library.Peaks were searched manually with a forward and reverse match of at least 800/1000 used as the threshold, otherwise the compounds were labeled as unknown.

## 3. Results & Discussion

Table 1 shows the compounds identified in the headspace above the plasmodial stage and the sporulated stage at two different temperatures 30$^{o}$C and 75$^{o}$C. A total of 115 compounds were identified from all the samples. A relatively small number of peaks were also measured but not identified (unknown compounds) and these are not includedin Table 1, although further work to identify them is ongoing. The analysis of the blank control samples yielded minimal interfering volatiles. Figure 1 shows typical chromatograms for each set of experiments. A total of 87 compounds were identified in the headspace above the plasmodial stage of the slime mould at 75$^{o}$C whereas this dropped to 79 compounds when extracted at 30$^{o}$C. Although undoubtedley 30$^{o}$C was more representative of volatiles released from the living organism, extraction at a higher temperature often enabled larger quantities of certain volatiles to be extracted which aided in the mass spectral identification. Far fewer compounds were identified in the post-sporulation stage of the slime mould with 50 compounds identified at 75$^{o}$C dropping to 37 compounds at 30$^{o}$C.

Figure 2. Shows the % abundance of certain classes of compounds found above the plasmodial and spoulating phases at the two extraction temperatures. The % abundance of compounds in the plasmodial phase at both extraction temperatures are almost identical. Therefore, we can be relatively confident that the higher extraction temperature is driving organic compounds into the headspace and not producing additional breakdown products and changing the chemical composition. If the case of the spores is considered then the profiles are quite similar apart from a lack of aromatic compounds in the extraction at 30$^{o}$C. We can postulate that these compounds are probably present at levels below the limit of detection for the method at this extraction temperature. Thus it highlights the benefits of using two extraction temperatures.

The most abundant classes of compounds identified in the headspace above the plasmodial stage were aromatics = aldehydes > ketones = terpenes = alcohols > hydrocarbons = esters = furans > volatile sulphur compounds > nitrogen containing compounds = acids. This compares to the post-sporulation stage where  ketones > aromatics > aldehydes. Then all other classes of compounds were present at relatively similar levels within the remaining classes. There were some differences in the numbers within each class extracted at different temperatures. However, in general the percentage composition of volatiles in each class was maintained especially for the plasmodial stage.

In the sporulated stage the number of aromatics, aldehydes, furans, alcohols and terpenes decreased markedly compared to the plasmodial phase. There was also a slight decrease in esters and hydrocarbons. Whereas, the number of ketones were maintained as were the nitrogen containing compounds and acids. There were 48 compounds identified above the plasmodial samples that were not present in the post-sporulation stage. Whereas, there were only 15 compounds present in the post-sporulation stage which were not identified above the plasmodial stage. Of the 48 compounds only found in the plasmodial stage aldehydes were the most abundant group with 11 not present in the post-sporulation stage. This was followed by aromatics (8), esters (5), terpenes (5), hydrocarbons (5), alcohols (5), furans (4), ketones (3) and volatile sulphur compounds (2). However, if this is calculated as a percentage of the total number of volatiles present in each class, then 80% of the furans are absent from the sporulating phase, 71% of the alkanes, 55% of the aldehydes and terpenes, 50% of the volatile sulphur compounds and esters, 45% of the alcohols, 38% of the aromatics and 17% of the ketones. Therefore, no acids or nitrogen containing compounds were absent from the post-sporulation stage. If the compounds missing from the plasmodial stage but present in the sporulating stage are considered then the most abundant group is ketones (6), followed by nitrogen containing compounds (4), esters (2), acids (2) and alcohols (1). This means that no alkanes, aromatics, terpenes, aldehydes or volatile sulphur containing compounds were found to be missing from the plasmodial stage, but present in the post-sporulation phase. In terms of the percentage of the total number of volatiles in each class 80% of the volatile nitrogen containing compounds were absent from the plasmodial stage, 50% of the acids, 35% of the ketones, 20% of the esters and 9% of the alcohols. Therefore, it would seem to be the case that the nitrogen containing compounds identified are a marker for the sporulating phase. It is known that nitric oxide synthase is induced during the sporulation of *Physarum polycephalum* [Golderer et al. 2001]. However, it is not clear whether other nitrogen containing compounds are produced in response to sporulation, to induce sporulation or as a reaction byproduct of other nitrogen containing compounds such as nitric oxide realeased at the point of sporulation. Conversely, furans and alkanes in particular seem to be associated with the plasmodial stage.

A number of VOCs associated with fungi were identified in the headspace above the plasmodial stage (and spores) of *P. polycephalum*. In particular 1-octen-3-ol is strongly associated with a number of fungi and is a semiochemical with an earthy, mushroom odor [Kaminiski *et al.* 1974]. Fungi produce 1-octen-3-ol via the enzymatic oxidation of linoleic acid [Chittara 2004]. Other fungal volatiles identified include 3-octanol, 2-methyl-1-propanol, 1-octen-3-one, and 3-octanone. Interestingly in previous work 3-octanone was identified as a compound which induced positive chemotaxis in the plasmodial stage of the slime mould *P. polycephalum* [de Lacy Costello and Adamatzky 2013]. 2-methyl-1-propanol was identified above both plasmodial stage and sporulating stage and is known to be derived from the breakdown of organic matter by bacteria [Atsumi *et al*. 2008]. It has also previously been identified as an attractant for fungivores [Morath *et al.* 2012]. 2-methyl propanoic acid was identified above the spores and has previously been identified as having antifungal properties [Morath *et al.* 2012]. It is also known to be produced by bacteria. In addition to this α-Muurolene was identified, this is a sesquiterpene and known fungal metabolite found in *Coprinopsis cinerea* [Agger *et al.* 2009].

The terpenes β-myrcene, β-farnesene and limonene were all identified in the headspace above the plasmodial stage of the slime mould. This is interesting as in previous work to assess the chemotaxis behavior of the slime mould all three of these terpene derivatives had induced a strong positive chemotactic response from the plasmodial stage [de Lacy Costello and Adamatzky 2013]. Indeed of all the 19 compounds tested ranging a wide variety of functional groups the sesquiterpene β-farnesene exhibited the strongest chemotactic response, followed by β-myrcene. These compounds have a relatively similar chemical structure being non cyclic terpene derivatives. However, we also identified benzaldehyde as a secretion of the slime mould and in chemotactic assays this was found to be a very strong inhibitor of chemotaxis. In common with many bio-active compounds the concentration may be important in controlling the response. The extraction technique used here has the ability to detect low vppb of certain compounds. In contrast chemotactic assays used higher concentrations initially. Without further work it would be difficult to establish the exact functionality of these secretions (if any). However, in other organisms such as fungi it is well known that VOCs act as signaling molecules (semiochemicals) that impart affects on other organisms within the same species, among species or more widely across kingdoms [Cottier and Mühlschlegel 2012]. Thus their secretions are known to act in defence of certain predators, but conversely other VOCs can act as attractants for spore dispersal etc. Thus fungal VOCs act as pheromones, allomones, kairomones etc. especially of certain insect species [Mburu *et al.* 2011]. It is likely that the VOCs identified above *P. polycephalum* have similar bi-directional functionality that has evolved within the ecosystems which it inhabits.

Another interesting compound identified in the headspace samples was geosmin, which is strongly associated with an earthy aroma [Gerber and Lechevalier 1965]. Indeed it is mainly responsible for the strong smell when rain falls after dry weather or after soil has been disturbed. This is because it is produced by a number of classes of soil dwelling microbes including cyanobacteria and actinobacteria. It was established that the actinobacteria *Streptomyces coelicolor* biosynthesizes geosmin by converting farnesyl diphosphate via a two step process using the bifunctional enzyme geosmin synthase [Jiang *et al.* 2006]. Bacterial DNA was identified during the ongoing genome project [Glockner *et al.* 2008] of *P. polycephalum*. The reasons for this could be ingestion as food (or transiently), surface contamination of the plasmodium or it is possible it lives in symbiosis with certain bacterial species –just as many other organisms do. Physarum has the ability to degrade lipopolysaccharides (LPS) from a variety of bacteria [Saddler et al. 1979]. Thus although the origin of certain volatiles is uncertain it is possible that they derive from an indirect source rather than directly from the metabolic processes of the slime mould. It should be noted that this is also the case for many VOCs attributed to humans and other organisms which are often derived from the interaction of metabolites with the associated microbiota and not directly from the host. Many of the volatiles identified in table 1. have also been identified as volatile metabolites from healthy human subjects [de Lacy Costello *et al.* 2013]. Physarum has been used as a model system for eukaryotic mitotic cycle, cell differentiation and motility [Dove and Rusch 1980]. In addition it is a model system for understanding the biochemical and physiological control of growth and development. Therefore, it is probably not surprising that Physarum shares common metabolites with a range of other eukaryotic organisms. Also as mentioned many eukaryotic organisms host various prokaryotic organisms which add to their metabolic profile. Physarum would also be subject to the same environmental exposure (as humans and other organisms) to exogenous volatile compounds when it is cultured artificially in the laboratory.

The sesquiterpene alcohol spathulenol was identified in the headspace above the slime mould. This compound makes up 7% of the oil of *Hypericum perforatum* (St Johns Wort) and 20% of the oil of *Nepeta species* (catnip) [Baser *et al.* 2000]. In previous work [Adamatzky 2011] chemotactic assays were performed using *Nepeta cataria* and it had a moderate chemoattractive effect on the slime mould. Many other somnipherous and psychoactive herbal based products have been shown to have an affect on the behavior of the slime mould *P. polycephalum*, although the specific chemicals responsible for the effect have not yet been identified. Also identified was β-Vatirenene which is a sesquiterpene that has been isolated from the roots and makes up 1% of the oil of N*ardostachys jatamansi* (Spikenard) a flowering plant of the Valerian family [Parveen *et al.* 2011]. *Valeriana officinalis* was found to be a very strong chemo-attractant for *P. polycephalum*, although the exact chemicals responsible for the effect were not identified [Adamatzky 2011]. Interestingly cadalene was identified in the headspace of the slime mould, this is a polycyclic aromatic hydrocarbon derived from sesquiterpenes and is ubiquitous in the oils of many higher plants [van Aarsen *et al.* 1992]. Other complex molecules identified included the ester Linalyl anthranilate which makes up 12% of the oil of lavender, which is known to have strong anti-bacterial properties [Adaszynska *et al.* 2012]. Methylisoquinolone was also identified which is a known bioactive compound that inhibits Cytochrome P450 enzymes [Rahnasto *et al*. 2005].

The aim of the work was to measure the volatiles directly from *P. polycephalum*. Therefore, samples were collected of the plasmodium growing across an inert substrate without media. In the same way spores were collected from *P. polycephalum* that had sporulated on an inert media. However, it should be noted that *P. polycephalum* had previously been cultured on non-nutrient agar with the addition of oat flakes. Therefore, as with all studies of this nature it is impossible to exclude the effects of nutrients, pathogenic microbes, commensal microbes etc. It is also difficult to replicate natural conditions/nutrients for the plasmodium.

This work has presented new data on the volatile metabolites extracted from the headspace above the plasmodial stage and sporulating stage of *P. polycephalum*. It was beyond the scope of the current study to investigate the function of these volatiles. However, this work could form the basis of further metabolomic and functional investigations of *P. polycephalum*. Future work could also incorporate these volatiles into simple chemotactic assays monitored against VOCs and more complex molecules already known to have an effect on *P. polycephalum*. The overarching aim of our work is to control the growth direction, speed and morphology of the plasmodial stage. This would hopefully enable directed transport of nanoparticles by the slime mould and the creation of functional materials.

## 4. Conclusion

This work presents new data on the volatile metabolites excreted into the headspace above the plasmodial stage and post-sporulation stage of *P. polycephalum*. In total 115 compounds were identified. These compounds spanned a diverse range of chemical classes. The number of compounds identified in the plasmodial stage was higher than the post-sporulation stage. There were also distinct differences in some chemical classes with nitrogen containing compounds more common after sporulation and furans more common in the plasmodial phase. There were also changes in other specific VOCs. Two extraction temperatures were used 75$^{o}$C and 30$^{o}$C. The rationale was to measure the slime mould in a living state with a non-destructive extraction technique. The higher temperature was used to increase the levels of volatiles extracted to provide better mass spectra. Actually there were not large differences in the VOC profiles when comparing the plasmodial phase extracted at the two temperatures. There were larger differences in the sporulated phase which probably just reflects the lower concentrations of volatiles in this phase *per se*.

A large number of the volatiles identified had been previously associated with fungal or bacterial metabolites most notably 1-octen-3-ol and geosmin. Interestingly a number of the identified volatiles including β-farnesene, β-myrcene, 3-octanone and limonene had previously been shown to exert a positive chemotactic response from the plasmodial stage of *P. polycephalum*. Future work could incorporate more of the volatiles identified into similar chemotactic assays. In addition more work needs to be carried out to establish the function of these VOCs within *P. polycephalum*. Future work would also investigate whether VOCs could be used to control the growth rate and morphology of *P. polycephalum*.

## 5. Acknowledgements

The authors would like to acknowledge the support of a Samsung GRO award.

**Table 1: volatiles detected from (*Physarum polycephalum*) plasmodial stage and the sporulating stage using SPME GCMS**

| | | Plasmodia | | Spores | |
|---|---|---|---|---|---|
| Volatiles | Retention Time | Extracted at (75℃ for 1h) | Extracted at ( 30℃ for 3h) | Extracted at (75℃ for 1h) | Extracted at (30℃ for 3h) |
| **Aldehydes** | | | | | |
| Acetaldehyde | 3.45 | + | + | + | + |
| 2-Propenal | 5.19 | + | | + | |
| Methacrolein | 5.92 | + | + | | |
| 2-methyl-Propanal | 6.81 | + | + | + | + |
| (E)-2-Butenal | 7.14 | + | | | |
| 3-methyl-Butanal | 9.73 | + | + | + | + |
| 2-methyl-Butanal | 9.97 | + | + | + | + |
| Pentanal | 11.12 | + | + | | |
| (E)-2-methyl-2-Butenal | 13.00 | + | + | | |
| Hexanal | 14.73 | + | + | | |
| 3-Methyl-2-butenal | 14.93 | + | | | |
| trans-2-Ethyl-2-hexenal | 17.39 | + | + | | |
| Heptanal | 18.73 | + | | | |
| (Z)-2-Heptenal | 21.68 | + | | | |
| Octanal | 22.90 | + | | | |
| Nonanal | 27.06 | + | + | + | + |
| (E)-2-Nonenal | 30.06 | + | + | | |
| Decanal | 31.11 | + | + | + | + |
| Dodecanal | 37.75 | + | | | |
| Pentadecanal | 47.96 | + | + | | |
| *Total aldehydes* | | **20** | **13** | **7** | **6** |
| **VSCs** | | | | | |
| Methanethiol | 3.69 | + | | + | |
| Dimethyl sulfide | 5.51 | + | + | | |
| 3-(methylthio)-Propanal | 20.10 | + | | | |
| Cyclohexyl isothiocyanate | 33.78 | + | + | + | + |
| *Total VSCs* | | **4** | **2** | **2** | **1** |
| **Acids** | | | | | |
| Acetic acid | 9.84 | + | + | + | + |
| 2-methyl-Propanoic acid | 14.69 | | | | + |
| 2-methyl-Butanoic acid | 18.16 | | | | + |
| 3-methyl- Butanoic acid | 17.93 | + | + | + | + |
| *Total Acids* | | **2** | **2** | **2** | **4** |
| **Alcohols** | | | | | |
| Ethanol | 4.80 | + | + | | |
| 2-Propanol | 5.64 | + | + | | + |
| 1-Propanol | 7.36 | | + | | |
| 2-methyl-1-Propanol | 9.32 | + | + | | + |
| 1-Pentanol | 12.78 | + | + | + | + |
| 2-methyl-1-Butanol | 12.90 | + | + | + | + |
| 2-Methyl-2-buten-1-ol | 14.24 | | + | | |
| 2-Methyl-3-pentanol | 15.89 | | | | + |
| 1-Hexanol | 17.72 | + | | | |
| 1-Octen-3-ol | 22.00 | + | + | + | |
| 3-Octanol | 22.53 | + | + | | |
| *Total Alcohols* | | **8** | **9** | **3** | **5** |
| **Furans** | | | | | |
| Furan | 5.02 | + | + | | |
| 2-methyl-Furan | 7.58 | + | + | | |
| Furfural | 17.20 | + | + | | |
| 2-pentyl-Furan | 21.43 | + | + | | |
| trans-tetrahydro-5-methyl-2-Furanmethanol | 21.56 | | + | + | + |
| *Total Furans* | | **4** | **5** | **1** | **1** |
| **Ketones** | | | | | |
| Acetone | 5.41 | + | + | + | + |
| 2,3-Butanedione | 7.89 | + | + | + | + |
| 2-Butanone | 8.10 | + | | + | + |
| 2-Pentanone | 10.87 | | | | + |
| 2,3-Pentanedione | 11.06 | | | | + |
| 3-Pentanone | 11.13 | | | + | |
| 3-hydroxy-2-Butanone | 12.64 | + | + | + | + |
| 2,3-Heptanedione | 16.00 | + | + | + | + |
| 2-Hydroxy-3-pentanone | 16.15 | | | | + |
| 4-methyl-3-Hexanone | 16.25 | | | + | |
| 5-Methyl-2-hexanone | 18.38 | | + | | |
| 1-Octen-3-one | 21.92 | | + | | |
| 3-Octanone | 22.16 | + | + | + | |
| Butyrolactone | 22.37 | | | | + |
| Benzyl methyl ketone | 29.69 | + | + | + | + |
| 4-(2,6,6-Trimethylcyclohexa-1,3-dienyl)but-3-en-2-one | 40.18 | + | + | + | + |
| 3-Nonanone | 30.76 | + | + | | |
| *Total Ketones* | | **9** | **10** | **10** | **11** |
| **Hydrocarbons** | | | | | |
| 1,3-Cyclohexadiene | 9.63 | + | | | |
| Octane | 13.16 | + | + | + | |
| 1,3-Octadiene | 14.51 | + | + | | |
| (Z)-2-Hexene | 15.20 | | + | | |
| Undecane | 20.83 | + | + | + | |
| cis-1-Ethyl-4-Methylcyclohexane | 24.36 | | + | | |
| Tetradecane | 38.50 | + | + | + | + |
| *Total Hydrocarbons* | | **5** | **6** | **3** | **1** |
| **Terpenes and terpenoids** | | | | | |
| β-Myrcene | 21.07 | + | + | | |
| Limonene | 23.03 | + | + | + | |
| Carvomenthol | 29.12 | + | + | | |
| β-Farnesene | 38.31 | + | + | | |
| Geosmin | 38.69 | + | + | + | + |
| β-Vatirenene | 39.58 | + | + | + | |
| α-Muurolene | 39.69 | + | + | | |
| Spathulenol | 40.34 | + | + | + | |
| Cadalene | 45.06 | + | + | | |
| *Total terpenes* | | **9** | **9** | **4** | **1** |
| **Esters** | | | | | |
| Formic acid, ethenyl ester | 6.62 | + | | | |
| Ethyl Acetate | 8.16 | | + | | |
| 2-methylene Butanoic acid methyl ester | 16.98 | | + | | |
| 2-Pentyl acetate | 17.10 | | + | | |
| Methyl α-methylacetoacetate | 17.15 | | | | + |
| 2-hydroxy-4-methyl-Pentanoic acid methyl ester | 21.93 | | | | + |
| iso-Butyl tiglate | 26.02 | | | | + |
| Linalyl anthranilate | 26.90 | + | + | + | |
| 2-hydroxy-Benzoic acid methyl ester | 31.76 | + | + | + | |
| 2-methyl-2-ethyl-1-propyl-1 Propanoic acid 3-propanediyl ester | 42.32 | + | + | | |
| *Total Esters* | | **4** | **6** | **2** | **3** |
| **Nitrogen containing compounds** | | | | | |
| Acetonitrile | 5.88 | | | | + |
| Pyrrole | 14.70 | | | + | |
| Hexanenitrile | 18.96 | + | | | |
| 2,5-dimethyl-Pyrazine | 19.04 | | | | + |
| trimethyl-Pyrazine | 22.68 | | | | + |
| 4,4a,5,6,7,8,9,9a-octahydro- | 32.01 | + | + | + | |
| 10,10-dimethyl-1,4-Methano-1H-cyclohepta[d]pyridazine | | | | | |
| 1-Methylisoquinoline | 36.42 | + | | + | |
| *Total Nitrogen containing compounds* | | **3** | **1** | **3** | **3** |
| **Aromatic and Benzenoids** | | | | | |
| Benzene | 9.54 | + | + | + | |
| Styrene | 18.15 | + | + | + | |
| p-Xylene | 18.36 | + | | + | |
| o-Xylene | 18.62 | + | | | |
| Anisole | 19.45 | + | + | + | |
| Benzaldehyde | 22.36 | + | | + | |
| Phenol | 25.36 | + | + | + | + |
| Benzeneacetaldehyde | 25.98 | + | + | + | |
| 4-ethyl-3-methyl-Phenol | 27.52 | | + | | |
| 2-phenyl-2-Propanol | 27.64 | | + | | |
| 1-Methoxy-4-vinylbenzene | 29.55 | + | + | | |
| 4-Ethylphenol | 32.50 | + | + | | |
| 4-Ethyl-1,2-dimethoxybenzene | 35.89 | + | + | + | |
| 4-ethenyl-1,2-dimethoxy benzene | 37.26 | + | + | + | |
| 1,2,4-Trimethoxybenzene | 37.47 | + | + | + | |
| 1,2-Dimethoxy-4-(1-methoxyethenyl) benzene | 40.78 | + | + | | |
| 1,2,3,4- | 41.42 | + | | | |
| Tetramethoxybenzene | | | | | |
| p-Acetylanisole | 37.95 | + | | | |
| 2-Methoxy-4-ethylphenol | 35.36 | + | + | + | |
| Benzophenone | 44.95 | + | + | + | |
| 2-tert-Butyl-4-(dimethylbenzyl)phenol | 46.33 | + | + | + | |
| *Total Aromatic and Benzenoids* | | 19 | 16 | 13 | 1 |
| *Over all total number of compounds* | | **87** | **79** | **50** | **37** |

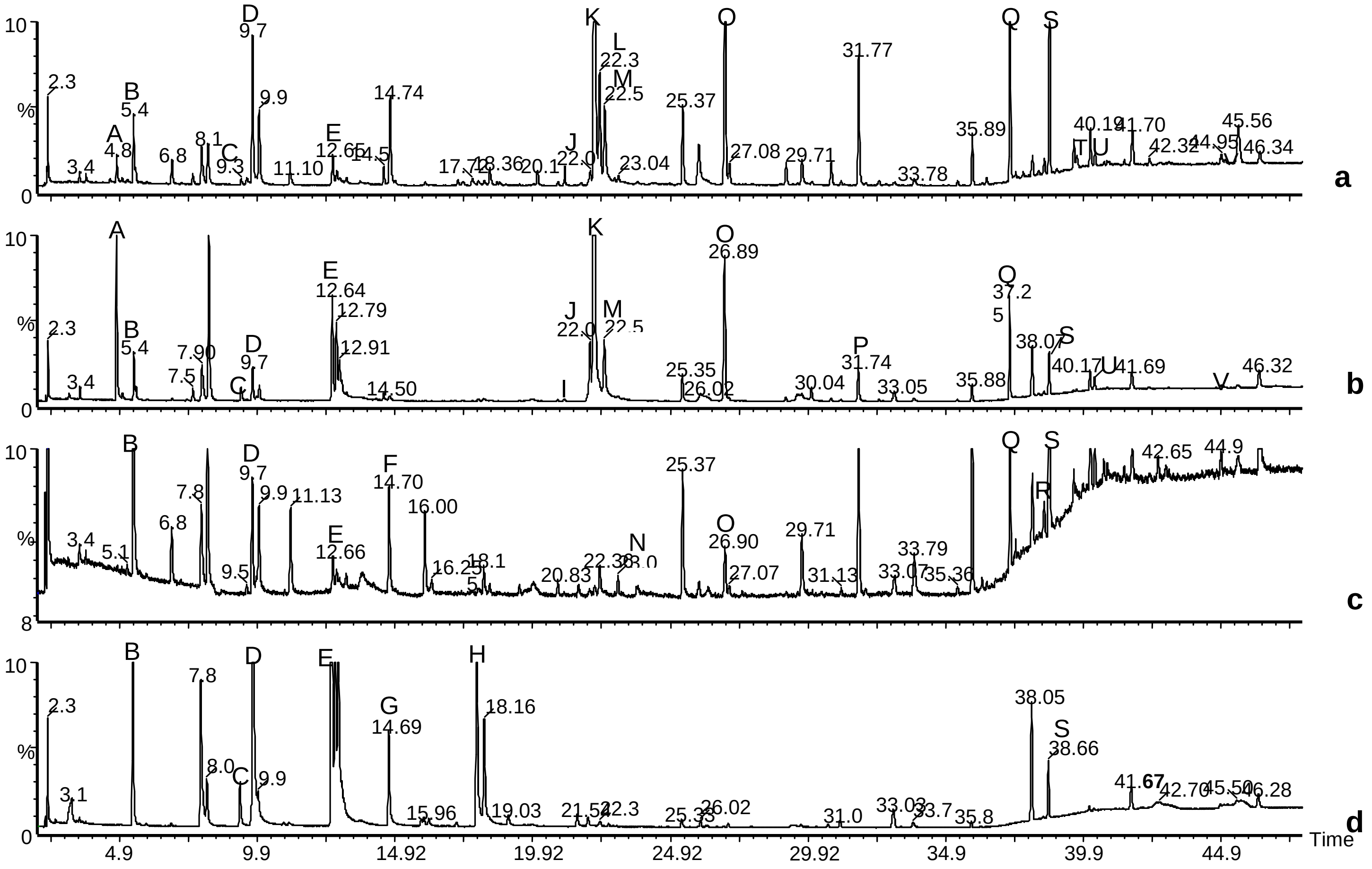

**Figure 1**. Chromatograms showing the volatiles detected in the headspace above the plasmodial stage (a: slime mould extracted at 75°C for 1h, b: slime mould extracted at 35°C for 3h) and post sporulation phase (c: slime mould spores extracted at 75°C for 1h, d: slime mould spores extracted at 35°C for 3h) of *Physarum polycephalum*. Selected volatiles are identified on the chromatograms A: Ethanol, B: Acetone, C: 2-methyl-1-Propanol, D: 3-methyl-Butanal, E: 3-hydroxy-2-Butanone, F: Pyrrole, G: 2-methyl-Propanoic acid, H: Methyl α-methylacetoacetate, I: β-Myrcene, J: 1-Octen-3-ol, K: 3-Octanone, L: Benzaldehyde, M: 3-Octanol, N: Limonene, O: Linalyl anthranilate, P: 2-hydroxy-Benzoic acid methyl ester, Q: 4-ethenyl-1,2-dimethoxy benzene, R: β-Farnesene, S: Geosmin, T: α-Muurolene, U: Spathulenol, V: Cadalene.

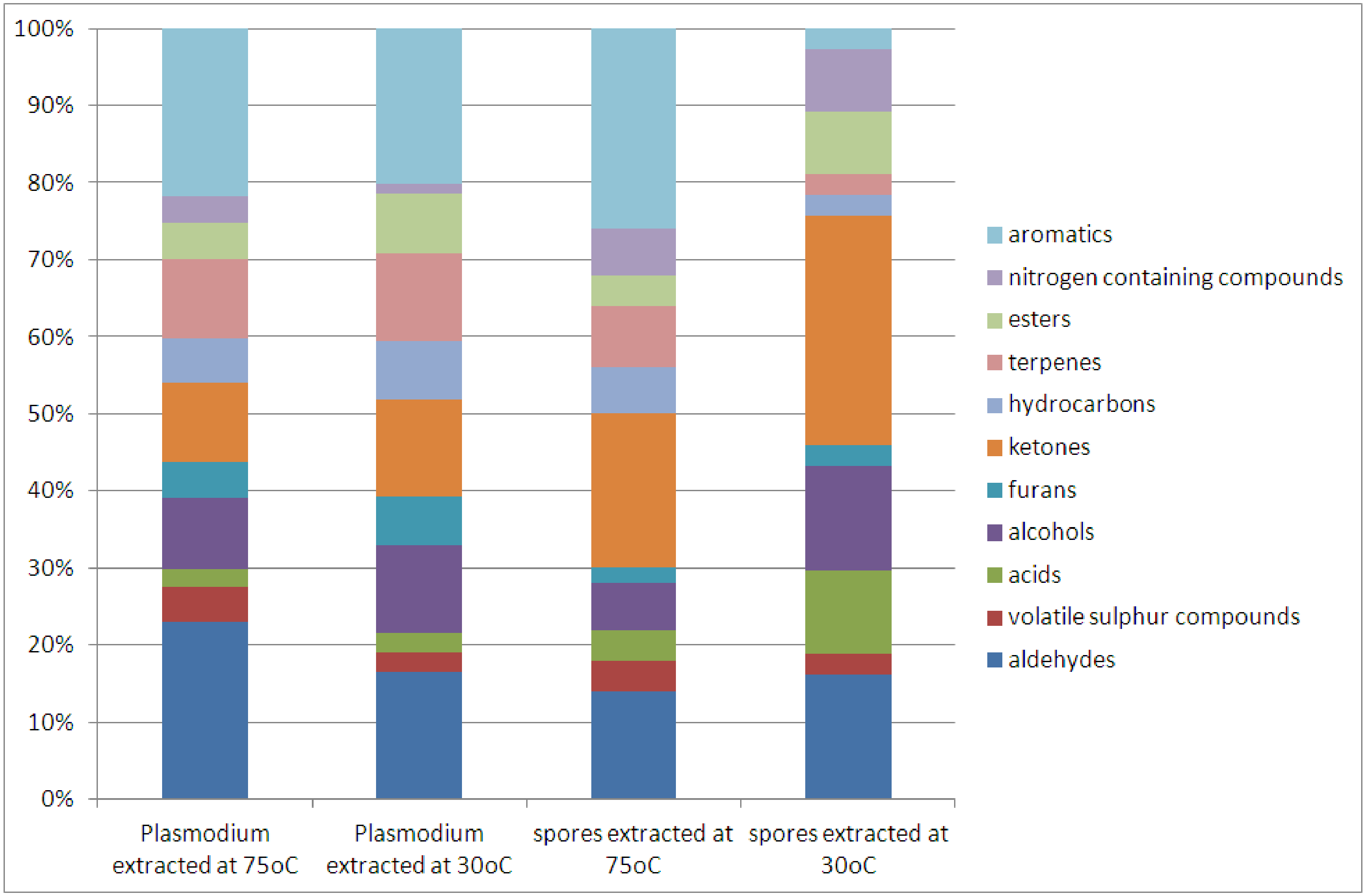

**Figure 2**. Shows the % abundance of certain classes of volatile organic compounds extracted from the headspace above the plasmodial stage and sporulating stage of *Physarum polycephalum*.